\documentstyle[epsfig,sprocl]{article}

\input{epsf.tex}
\font \fwork = cmssqi8
\font \fcoll = cmr9

\begin{document}
\hspace*{-18pt}{\epsfxsize=80pt  \epsfbox{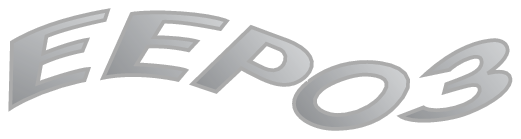}} 
\vspace*{-16pt}
\rightline{\fwork LPSC Grenoble, France, October 14-17, 2003}
\vspace*{15pt}

\def\micro{\mu}
\def\deg{^\circ}
\def\gtorder{\mathrel{\raise.3ex\hbox{$>$}\mkern-14mu
 \lower0.6ex\hbox{$\sim$}}}
\def\ltorder{\mathrel{\raise.3ex\hbox{$<$}\mkern-14mu
 \lower0.6ex\hbox{$\sim$}}}
\def\mugegm{\mu_p G_E / G_M}
\def\gegm{G_E / G_M}
\def\ge{G_E}
\def\gm{G_M}
\def\etal{\textit{et al.}}


\title{\Large\bf New measurement of $\gegm$ for the proton.}

\author{J. Arrington, \\ 
\fcoll{for the JLab E01-001 Collaboration}}

\address{{\it Physics Division, Argonne National Laboratory, Argonne, IL 60439}}

\maketitle

\begin{abstract}

Recent polarization transfer measurements of the proton form factors disagree
with previous Rosenbluth extractions.  In addition to providing insight into
the proton structure, the form factors also provide crucial input for the
analysis of quasielastic experiments, and this discrepancy could have a
significant impact on such experiments at moderate-to-large $Q^2$ values. In
2002, a new ``SuperRosenbluth'' measurement was performed to help understand
this origin of this discrepancy, as well its implications on other
experiments. Preliminary results agree with previous Rosenbluth extractions,
indicating that the discrepancy must be related to some fundamental flaw in
either the Rosenbluth or polarization transfer technique.

\end{abstract}

\section{Introduction}

The proton electromagnetic form factors, $G_E(Q^2)$ and $G_M(Q^2)$, provide
important information on the substructure of the proton.  Several decades
worth of elastic electron-proton scattering measurements have provided
measurements of the proton form factors via the Rosenbluth separation
technique. Up to several GeV$^2$, both $\ge$ and $\gm$ approximately follow
the dipole form, and hence $\mugegm \approx 1$. However, at large $Q^2$
values, the cross section is dominated by $\gm$, and so extractions of $\ge$
are of limited precision.  Because of this, data above 2~GeV$^2$ are quite
sensitive to systematic uncertainties in the cross section measurements, and
there are no direct extractions of $\ge$ above $Q^2=9$~GeV$^2$.

The polarization transfer technique provides a direct measurement of $\gegm$,
with significantly reduced systematic uncertainties at large $Q^2$ values.
Recent polarization measurements indicate that $\ge$ falls more rapidly than
$\gm$ as $Q^2$ increases~\cite{jones00,gayou02}, with
\begin{equation}
\mu_p G_E(Q^2)/G_M(Q^2) = 1 - 0.13 (Q^2-0.04\mbox{GeV}^2).
\end{equation}
This significant falloff of the form factor ratio disagreements with the
Rosenbluth extractions of $\gegm$.  While the uncertainties in the Rosenbluth
extraction of $\ge$ grows rapidly with $Q^2$, they are not large enough to
allow for consistency with the polarization transfer
measurements~\cite{arrington03}.

Until it is resolved, the discrepancy between the form factors extracted
using these two techniques implies a large uncertainty in the proton form
factors.  Not only does this impact our knowledge of the proton structure,
it also has significant implications for many other experiments.  If either
the Rosenbluth or polarization transfer technique is incorrect, it may imply
similar problems for other experiments using these techniques.

The problem could also lie in the cross section data that goes into the
Rosenbluth extractions. The cross section data would have to have a significant
$\varepsilon$-dependent error, 5--8\% for $Q^2>2$~GeV$^2$, to explain the
discrepancy~\cite{arrington03,guichon03}. Because the elastic cross section
has routinely been used as a check on the normalization of electron scattering
experiments, this may lead to questions about several other experimental
results.  For quasielastic $A(e,e'p)$ measurements, precise information on the
elastic electron-proton scattering is even more important, as it is assumed
that the electron-proton cross section is well known when extracting
information on nuclear structure.

The cross sections one obtains from polarization transfer fits to the form
factors can differ significantly from those using Rosenbluth form
factors~\cite{arrington03b}. Fits that include the polarization transfer data
yield cross sections that are a few percent larger than the measured cross
sections at low $\varepsilon$, and a few percent smaller at large $\varepsilon$.
Combining inconsistent form factors can yield even larger discrepancies: 
combining $\gm$ from Rosenbluth measurements with $\gegm$ from polarization
transfer yields cross sections that are 5--10\% too low at large $\varepsilon$
for $0.5 < Q^2 < 10$~GeV$^2$.  It is crucial that we understand the
discrepancy, not only to be sure of the proton form factors, but also because
precise knowledge of the cross section is important to many electron
scattering experiments.

Because the difference in the form factors has the largest impact on the
$\varepsilon$-dependence of the cross sections, the uncertainty is
significantly enhanced for experiments which examine the longitudinal and
transverse response in quasielastic scattering.  JLab experiment
E91-013~\cite{dutta03} performed a Rosenbluth separation of the response
function for carbon at $Q^2=0.6$ and 1.8~GeV$^2$.  At low $Q^2$, they found an
excess of strength in the transverse response function, which they attribute
to many-body currents.  At $Q^2=1.8$~GeV$^2$, they find identical transverse
and longitudinal response functions if they use Rosenbluth form factors in
their analysis, but find that the longitudinal response is $\approx$60\% larger
than the transverse if they assume form factors extracted from polarization
transfer measurements. Clearly, these data cannot be properly interpreted
unless we know the appropriate form factors to use in such an analysis.

\section{JLab experiment E01-001}

The goal of JLab experiment E01-001 was to check the consistency of Rosenbluth
and polarization transfer measurements by performing a high-precision
Rosenbluth extraction of $\gegm$, with significantly reduced statistical and
systematic uncertainties compared to previous Rosenbluth separations.
Normally, one measures the elastic cross section at fixed $Q^2$, but varies
$\varepsilon$ by simultaneously varying the beam energy and electron
scattering angle.  Any corrections that depend on the beam energy or the
scattering angle are translated into $\varepsilon$-dependent corrections,
which can potentially change the extracted form factors. Varying $\varepsilon$
also implies a significant variation in the energy of the detected electron
the the elastic cross section, which can vary by two or more orders of
magnitude over the $\varepsilon$ range measured. Thus, any momentum-dependent
or rate-dependent corrections are also $\varepsilon$-dependent, and further
increase the uncertainty in the extracted form factors.

The main improvements in E01-001 come from detecting the struck proton, rather
than the scattered electron. In this case, the beam energy and proton angles
change, but the proton momentum is fixed and the cross section is nearly
constant, so there are no momentum-dependent corrections in the
$\varepsilon$-dependence at fixed $Q^2$, and the rate-dependent corrections
are many times smaller.  The large rate dependence when detecting electrons
also means that the low $\varepsilon$ measurements, corresponding to backward
angle electrons, are typically limited by lack of statistics.  
The minimum cross section is an order of magnitude larger for proton detection
then electron detection in our kinematics.  In addition, because the cross
section is nearly constant, measurements can be made with fixed beam currents,
which reduces the uncertainties related to possible current-dependence in the
beam current monitoring and target density fluctuations caused by local beam
heating.  Finally, detecting the proton reduces the uncertainties related to
knowledge of the beam energy and scattering angle, as well as reducing the
size and $\varepsilon$-dependence of the radiative corrections that must be
applied.

While there are corrections that are larger when detecting protons, they
generally have little effect on the extraction of $\gegm$. Proton absorption
and the use of acceptance-defining software cuts lead to significant
(few percent) uncertainties in the absolute cross section, but the corrections
are identical for all $\varepsilon$ values at a fixed $Q^2$, and therefore do
no lead to any additional uncertainty in the extraction of $\gegm$.
Backgrounds due to inelastic processes are also larger than for measurements
where the electron is detected. These are currently the dominant sources of
uncertainty in the measurement, but the overall systematic uncertainty in the
extracted form factor ratio, $\gegm$, is still significantly better than any
previous Rosenbluth separations in this $Q^2$ range.

\begin{figure}[htb]
\centerline{\epsfig{file=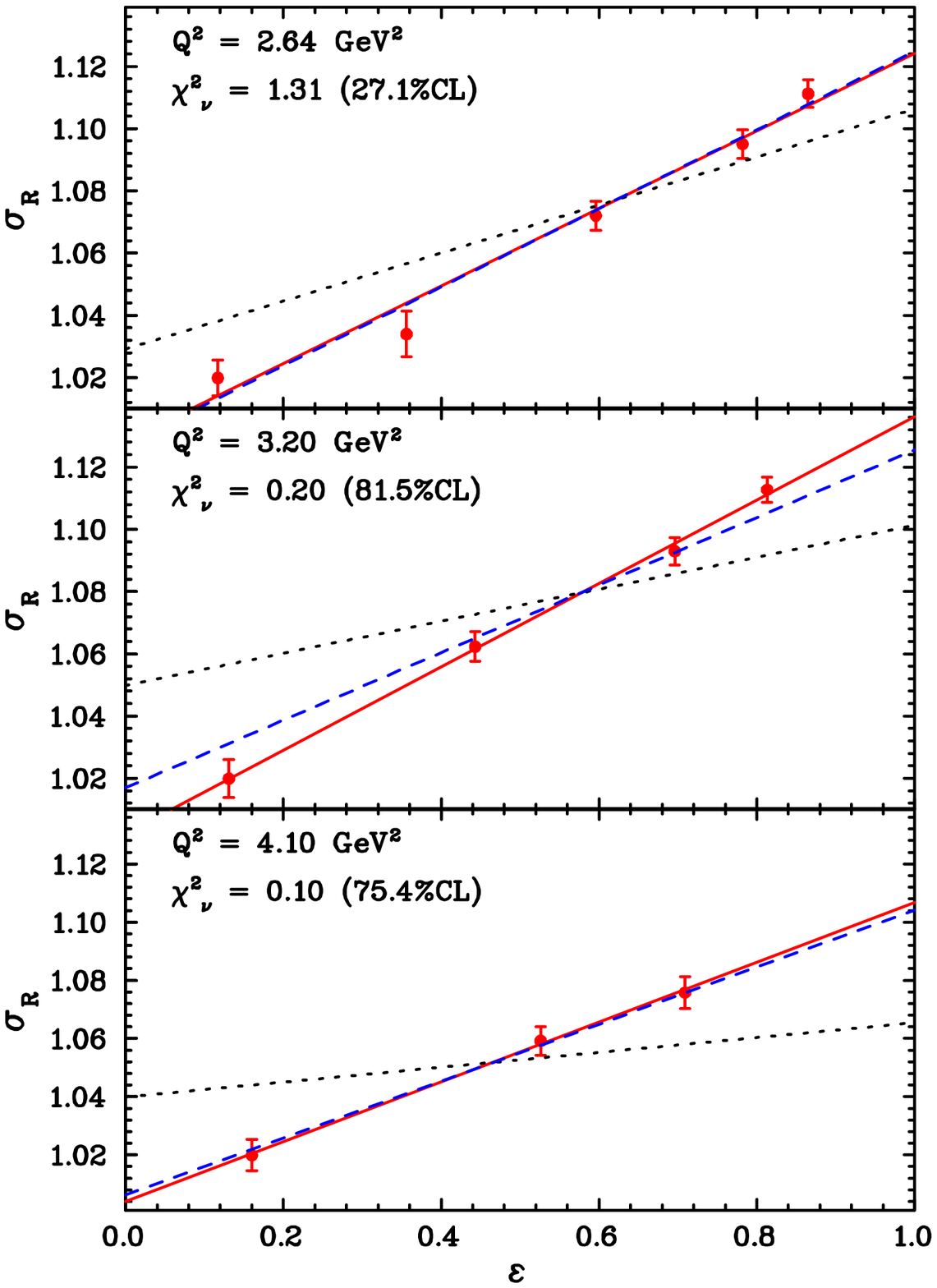,width=8.0cm,
height=9.0cm}}
\caption\protect{{The reduced cross section plotted as a function of
$\varepsilon$.  The solid line is the best linear fit to the data.  The black
dotted line indicates the expected slope as determined from polarization
transfer experiments~\cite{gayou02}, while the blue dashed line is the best
slope based of the global analysis of previous Rosenbluth
measurements~\cite{arrington03b}.  The new results confirm previous Rosenbluth
extractions (within the uncertainties on the global analysis), and disagree
with the polarization transfer results.}}
\label{fig:rosenbluth}
\end{figure}

Figure~\ref{fig:rosenbluth} shows the $\varepsilon$-dependence of the
reduced cross sections, along with a fit to the slope, which is proportional
to $(G_E/G_M)^2$.  The figure also shows the slope predicted by polarization
transfer measurements~\cite{gayou02} (dotted line) and the global Rosenbluth
analysis~\cite{arrington03b} (dashed line).  The uncertainties shown include
only the statistical and totally uncorrelated point-to-point systematic
uncertainties, so that the calculated $\chi^2$ values and confidence levels
are meaningful.  While the $\chi^2$ values are quite small for the large $Q^2$
points, there are only one or two degrees of freedom in the fits, and so the
confidence level is never unreasonably high or low. Additional
systematic uncertainties, not included in Fig.~\ref{fig:rosenbluth}, will also
contribute to the uncertainty to the extracted slope, and thus the value of
$\gegm$.

Clearly, the results are in much better agreement with previous Rosenbluth
extractions of $\gegm$ than with the results obtained by polarization transfer,
with significantly reduced systematic uncertainties compared to previous
measurements.  This result essentially rules out the possibility that the
disagreement is due to underestimation of the corrections or uncertainties due
to momentum- or rate-dependent systematic uncertainties in previous Rosenbluth
separation measurements.

\section{Interpretation}

One possible explanation for the discrepancy is a problem with the data or
analysis of the polarization transfer measurements.  However, the systematic
uncertainties of the polarization transfer experiment have been studied in
detail~\cite{punjabi03}, and there is no indication of any uncertainties large
enough to explain the discrepancy.

The other possibility is that the discrepancy is related to a more fundamental
problem with the Rosenbluth or polarization transfer techniques.  So far, most
work has focussed on the possibility that two-photon exchange corrections
may lead to an $\varepsilon$-dependent correction to the cross section, which
is not currently being taken into account. Such two-photon corrections were
generally estimated to be small, and this appeared to be confirmed by
measurements comparing positron and electron scattering.  However, recent
calculations of the two-photon corrections~\cite{blunden03}, and a
reexamination of the positron data~\cite{arrington03d}, indicate that
two-photon corrections may in fact play an important role in this kinematic
region.

If the discrepancy is explained by two-photon effects, then these need be be
understood in detail before we can reliably extract precise form factors. The
effect of two-photon exchange terms on the polarization transfer extraction
are expected to be relatively small, but this must be demonstrated
quantitatively.  In addition, since the polarization transfer extracts only
the ratio $\gegm$, the cross sections are still necessary to extract $\ge$
and $\gm$ separately, and so these corrections to the cross section must be
well understood.

While these data do not provide a full explanation of the discrepancy, 
excluding experimental error in the previous cross section measurements is
extremely important in determining the impact of this discrepancy on other
measurements. Since the discrepancy is not a result of errors in the previous
cross section measurements, the form factors extracted from a Rosenbluth
separation (a direct fit to the cross section data) will provide a good
parameterization of the measured elastic electron-proton cross section, even
if the form factors do not correspond to the `true' form factors of the
proton.  These form factors are therefore the appropriate ones to use when
using elastic scattering to check experimental normalizations.  They are also
appropriate as input to the analysis of quasielastic experiment, unless the
missing corrections are different for $A(e,e'p)$ and $p(e,e'p)$. Conversely,
the polarization form factors \textit{will not} yield the correct
electron-proton cross section, and will give incorrect results if used as
input to the analysis of experiments that rely on knowledge of the elastic
electron-proton cross section.

\section{Conclusions}

Preliminary result from a high-precision Rosenbluth separation measurement of
$\gegm$ from Jefferson Lab E01-001 are in good agreement with previous
Rosenbluth extractions.  They clearly disagree with polarization transfer
measurements of similar precision.  This shows that the discrepancy is
more than a simple experimental problem with previous Rosenbluth extractions. 
While the new results from this experiment do not provide a full explanation
of the discrepancy, excluding experimental error in the previous cross section
measurements is extremely important in determining the impact of this
discrepancy on other measurements.  If the discrepancy is not a result of
errors in the previous cross section measurements, then the form factors
extracted from a Rosenbluth separation will provide a good parameterization of
the measured elastic electron-proton cross section, even though they may not
represent the underlying structure of the proton.  Similarly, form factors
extracted from the polarization transfer measurements may represent the true
form factors, but do no provide a good parameterization of the elastic cross
section.

It is quite possible that two-photon exchange corrections, only partially
taken into account in standard radiative correction prescriptions, may
be the explanation for the discrepancy.  However, until the details are
understood, we cannot be certain of the proton form factors. The two-photon
contributions to the polarization transfer reactions are assumed to be small,
but may not be negligible~\cite{guichon03}, and even if we assume that the
discrepancy is due \textit{entirely} to corrections to the cross section,
there are still small uncertainties in the form factors extracted from a
combined analysis if we do not fully understand these missing
correction~\cite{arrington03b}. While further theoretical and experimental
work is clearly required to fully understand the importance of these two-photon
contributions, the confirmation of previous Rosenbluth measurements answers an
important experimental question, directly relevant to this workshop:  What are
the correct form factors to use as input in the analysis of quasielastic data?

\smallskip

\textit{This work was supported in part by the U. S. Department of Energy,
Nuclear Physics Division, under contract W-31-109-ENG-38.}

\end{document}